\documentclass[fleqn]{2023SCGE}
\usepackage{hyperref}%PDF??????
\usepackage{tabularx}
\usepackage{longtable}
\usepackage{extarrows}
\usepackage{supertabular}
\usepackage{threeparttablex}
\usepackage{graphicx}% Include figure files
\usepackage{dcolumn}% Align table columns on decimal point
\usepackage{bm}% bold math
\usepackage{amsmath}
\usepackage{overpic}
\usepackage{booktabs}%
\usepackage{makecell}
\usepackage{enumerate}
\usepackage{multirow}
\usepackage{rotating}[figuresright]
\usepackage{booktabs}
\usepackage{lineno} 
\begin{document}
% \linenumbers

\ensubject{subject}

%%%%%%%%%%%%%%%%%%%%%%%%%%%%%%%%%%%%%%%%%%%%%%%%%%%%%%%
%%% Authors do not modify the information below
%%% ????????????????
%%% ??????????, ????????????{}, ???????????????????
%Letter to the Editor??Article%??????
\ArticleType{Article}%??Article
\SpecialTopic{SPECIAL TOPIC: }%???????
\Year{*}
\Month{*}
\Vol{*}
\No{*}
\DOI{??}
\ArtNo{000000}
\ReceiveDate{****}
\AcceptDate{****}
%\OnlineDate{January 1, 2016}

\title{GECAM Discovery of Peculiar Oscillating Particle Precipitation Events}

\author[1,2]{Chenwei Wang}{}
\author[1]{Shaolin Xiong}{{xiongsl@ihep.ac.cn}}
\author[3]{Yi Zhao}{yizhao@dzu.edu.cn}
\author[4]{Wei Xu}{{wei.xu@whu.edu.cn}}
\author[5]{Gaopeng Lu}{{gplu@ustc.edu.cn}}
\author[6]{\\Xuzhi Zhou}{}
\author[7]{Xiaocheng Guo}{}
\author[7]{Wenya Li}{}
\author[7]{Xiaochao Yang}{}
\author[7]{Qinghe Zhang}{}
\author[1]{\\Xinqiao Li}{}
\author[8]{Zhenxia Zhang}{}
\author[1]{Zhenghua An}{}
\author[9]{Ce Cai}{}
\author[1,2]{Peiyi Feng}{}
\author[1]{\\Yue Huang}{}
\author[1]{Min Gao}{}
\author[1]{Ke Gong}{}
\author[1]{Dongya Guo}{}
\author[1,10]{Haoxuan Guo}{}
\author[1]{Bing Li}{}
\author[1]{\\Xiaobo Li}{}
\author[1]{Yaqing Liu}{}
\author[1,2]{Jiacong Liu}{}
\author[1]{Xiaojing Liu}{}
\author[1]{\\Xiang Ma}{}
\author[1]{Wenxi Peng}{}
\author[1]{Rui Qiao}{}
\author[1,11]{Yangzhao Ren}{}
\author[1]{Liming Song}{}
\author[1,2]{Wenjun Tan}{}
\author[1]{\\Jin Wang}{}
\author[1]{Jinzhou Wang}{}
\author[1]{Ping Wang}{}
\author[1,2]{Yue Wang}{}
\author[1]{Xiangyang Wen}{}
\author[1,12]{\\Shuo Xiao}{}
\author[1,13]{Shenglun Xie}{}
\author[1]{Yanbing Xu}{}
\author[1,2]{Wangchen Xue}{}
\author[1]{Sheng Yang}{}
\author[1,14]{\\Qibin Yi}{}
\author[1,2]{Zhenghang Yu}{}
\author[1]{Dali Zhang}{}
\author[1]{Fan Zhang}{}
\author[1,2]{Jinpeng Zhang}{}
\author[1,15]{Peng Zhang}{}
\author[1,16]{\\Wenlong Zhang}{}
\author[1,2]{Yanqiu Zhang}{}
\author[1,2]{Shuangnan Zhang}{}
\author[1,2]{Zhen Zhang}{}
\author[1]{Haisheng Zhao}{}
\author[1]{\\Xiaoyun Zhao}{}
\author[1,2]{Chao Zheng}{}
\author[1]{Shijie Zheng}{}

\address[1]{~~Key Laboratory of Particle Astrophysics, Institute of High Energy Physics, Chinese Academy of Sciences, Beijing 100049, Beijing, China}
\address[2]{~~University of Chinese Academy of Sciences, Beijing 100049, Beijing, China}
\address[3]{~~School of Computer and Information, Dezhou University, Dezhou 253023, Shandong, China}   
\address[4]{~~Electronic Information School, Wuhan University, Wuhan 430072, Hubei, China}   
\address[5]{~~School of Earth and Space Sciences, University of Science and Technology of China, Hefei 230026, Anhui, China}
\address[6]{~~School of Earth and Space Sciences, Peking University, Beijing, China}
\address[7]{~~State Key Laboratory of Space Weather, National Space Science Center, Chinese Academy of Sciences, Beijing, China}
\address[8]{~~National Institute of Natural Hazards, Ministry of Emergency Management of China, Beijing 100085, China}
\address[9]{~~College of Physics, Hebei Normal University, Shijiazhuang, Hebei, 050024, China} 
\address[10]{~~Department of Nuclear Science and Technology, School of Energy and Power Engineering, Xi'an Jiaotong University, Xi'an, China} 
\address[11]{~~School of Physical Science and Technology, Southwest Jiaotong University, Chengdu 611756, Sichuan, China}
\address[12]{~~School of Physics and Electronic Science, Guizhou Normal University, Guiyang, 550001, China}   
\address[13]{~~Institute of Astrophysics, Central China Normal University, Wuhan 430079, HuBei, China}
\address[14]{~~School of Physics and Optoelectronics, Xiangtan University, Xiangtan 411105, Hunan, China}
\address[15]{~~College of Electronic and Information Engineering, Tongji University, Shanghai 201804, China}
\address[16]{~~School of Physics and Physical Engineering, Qufu Normal University, Qufu, Shandong 273165, China}

% \AuthorMark{Wang C W}%\authorcr????????

\AuthorCitation{Wang C W, Xiong S L, Zhao Y, et al}

\abstract{

%大背景
Charged particle precipitation typically manifests as a gradual increase and decrease of flux observed by space detectors. Cases with rapidly flux variation are very rare. Periodic events are even more extraordinary.
These oscillating particle precipitation (OPP) events are usually attributed to the bounce motion of electrons, which are induced by lightning. Owing to the observation limitations, there has been debate regarding whether these oscillations originate from temporal flux evolution or spatial structure evolution.
%小背景
Here we report three peculiar charged particle precipitation events detected by GECAM during a geomagnetic storm on March 21, 2024, with two exhibiting significant periodicity. These events were observed around the same region during three consecutive orbits.
%概括结果
Through comprehensive temporal and spectral analyses, we revealed that one of the OPP events exhibited a transition in spectral lag of mini-pulses, shifting from ``softer-earlier" to ``softer-later" while showing no significant time evolution in overall frequency characteristics. And there is no association found between these two OPP events and lightning activity. 
%总结
Several possible scenarios are discussed to explain these charged particles with a life time of more than 3.5 hours, but the nature of these three events remains an enigma. 
We suggest that these GECAM-detected OPP events may represent a new type of particle precipitation event or a peculiar Lightning-induced Electron Precipitations (LEPs). 
}



\maketitle
\begin{multicols}{2}

% \Authorfootnote
\section{Introduction}\label{Sec_Int}
Charged particle precipitation, which is often observed during geomagnetic storms and substorms, plays an important role in space weather and Earth's atmospheric dynamics \cite{Osc_Bortnik2008}. The energetic precipitating charged particles, ranging from eV to MeV, significantly affect ionospheric chemistry and the propagation of electromagnetic waves, often leading to disruptions in communication systems and severe threats to the safety of spacecraft. Thus the detection of charged particle precipitation events is essential to understanding their generation and propagation processes, which is crucial to preventing their impact on the development and utilization of space.

One of the most important ways to accelerate and scatter charged particles is the various magnetohydrodynamic waves \cite{GRL_Thorne2010, GRL_Mozer2016}, while another important process is the reconnection of the magnetic field \cite{JGR_Hesse2020}. And both of them will accelerate charged particles, change the distribution of pitch angle, and scatter charged particles into the loss cone and precipitate, resulting in an increase in the flux of charged particles in the Low Earth Orbit (LEO) locally. 

Some charged particle precipitation events exhibit an oscillatory behavior, which is generally explained as being driven by the resonance with magnetohydrodynamic waves, including Electromagnetic Ion Cyclotron wave (EMIC), Very Low Frequency wave (VLF), Ultra Low Frequency wave (ULF), and so on, with periods ranging from seconds to days \cite{SCE_Zong2008,GRL_Foster2015,NC_Liu2022,NC_Liu2023,JGR_Li2024}. The spectral and temporal properties of these events provide hints to the underlying physical processes. For instance, the chorus wave typically leads to a gap in the power spectral density (PSD) at the frequency of one‐half the electron gyrofrequency \cite{GRL_Thorne2010, NC_Li2019}.

However, the discovery of sub-second periodic oscillations in high-energy particles is highly unexpected, if taking the corresponding resonance conditions into consideration \cite{ASTP_Millan2007,SCE_Zong2008}. Moreover, as the majority of space missions have mainly focused on changes in the space environment over longer time scales, specifically designed and optimized capabilities for high temporal resolution have been lacking. All these lead to an obstacle in our understanding of oscillating particle precipitation (OPP) events. 

This paper presents the high resolution detection of a pair of peculiar charged particle precipitation events, along with another associated event, which is regarded as a ``precursor", detected by the Gravitational-wave high-energy Electromagnetic Counterpart All-sky Monitor (GECAM). These two events exhibited significant sub-second periodicity in the keV to MeV energy range, which is exceptionally rare in space physics.

An overview of GECAM and these charged particle precipitation events is provided in Section \ref{Sec_Obs}, with a comprehensive analysis and discussion presented in Section \ref{Sec_Dis}. 
Finally, we conclude with insights obtained from the high resolution observations of OPP events in Section \ref{Sec_Con}.

\section{Instrument and Observation}\label{Sec_Obs}
GECAM-B is one instrument of the GECAM series telescopes, which is a constellation with four X-ray and gamma-ray all-sky space telescopes, including GECAM-A, GECAM-B \cite{GEC_INS_Li2022}, GECAM-C \cite{HEBS_INS_Zhang2023} and GECAM-D \cite{GTM_INS_wang2024}. 
GECAM-B was launched in December 2020 and is operating in low earth orbit (LEO) with an altitude of $\sim$600 km and 29$^{\circ}$ inclination angle \cite{GEC_INS_Han2020}. 

Two kinds of scientific payloads are equipped on GECAM-B, the gamma-ray detectors (GRDs) \cite{GEC_INS_An2022} and charged-particle detectors (CPDs) \cite{GEC_INS_Xv2021}. 
GRDs are highly sensitive to both photons and charged particles, while CPDs are designed to be only sensitive to charged particles and much less sensitive to photons, which makes GECAM able to distinguish whether a group of events is dominated by photons or charged particles. Although it is impossible to determine particle by particle whether it is a charged particle or a photon. 

There are 25 GRDs and 8 CPDs distributed in different directions on GECAM-B. Without considering the possibility of penetrating the satellite platform (which is almost impossible for charged particles with energies of about MeV), each GRD and CPD probe has a field of view (FOV) of 2$\pi$. 
Thus GECAM-B has the capacity to detect both X-ray, $\gamma$-ray and charged particles coming from almost all directions in the sky. Based on the signal from different detectors, GECAM-B can also estimate the incident direction of photons and the rough approximate spatial distribution of charged particles \cite{LOC_MCM_Liao2020,YIZ_LOC_GEC_2023}, which has been successfully applied in the detection of TGFs and TEBs \cite{YIZ_LOC_MTD_2023,YIZ_TGF_GEC_2023}.

When numerous particles incident into the detector in a short period with an extremely high flux, exceeding the processing capability of the onboard electronics, the data will be erroneously recorded as periodic signals due to instrument effects of transmission saturation. 
Many special efforts have ensured the reliability of GECAM data in detecting events with extremely high flux, including multi-channel readout, high-bandwidth transmission, and caching mechanisms \cite{GEC_INS_Liu2021}.
GECAM is also equipped with a comprehensive mechanism to check for transmission saturation. The system records the number of events processed by the electronics per second and the number of events transmitted, which should be the same in normal cases. Transmission saturation may have occurred only when there is an inconsistency between the scientific data and the engineering data. 
These delicately designed mechanisms of GECAM have proven effective in ensuring accurate measurement of high-flux events during the observation of the brightest gamma-ray bursts, GRB 221009A \cite{HXMT_GECAM_221009A,Zhang_09A} and the second brightest GRB, GRB 230307A \cite{GECAM_LEIA_07A,Yi_07A,Wang_07A}.

Additionally, as a telescope designed for time-domain astronomy, GECAM is delicately designed as one of the instruments that have the highest time resolution (0.1 $\mu$s) among all X-ray space telescopes ever flown, which is much higher than that of the vast majority of instruments designed for detecting charged particles in the space environment \cite{GEC_CAL_Xiao2022}. This provides a unique advantage for GECAM in detecting ultra-short timescale phenomena of charged particles in space, such as TGFs and TEBs \cite{YIZ_TGF_GEC_2023}, as well as the OPPs.

Some bump-shape signals, which are called space particle events, are observed by many wide-FOV gamma-ray monitors, including GECAM-B \cite{Huang2024alert}, \textit{Fermi}/GBM \cite{Briggs2007alert,Kienlin2020catalog} and so on. Most of the space particle events are actually particle precipitation caused by geomagnetic disturbance. 
The space particle events observed by GECAM-B can be roughly classified into two categories: local type and distant type. 
For local particle events, there is almost no overall directional motion of the population of charged particles, and all the detectors can detect incoming particles from almost all directions. 
For distant particle events, there is a clear overall directional distribution, meaning that the population of charged particles is moving towards a certain direction together, resulting in only a few detectors that have similar orientations having significant signals. 
GECAM can distinguish whether a particle event is local type or distant type based on the distribution of signals in different detectors and by checking if the same signal is present when revisiting again at the same position as well. 
The latter method (by revisit) has more limitations because the orbital period of GECAM-B is approximately 90 minutes, which requires local particle events to have a lifetime longer than at least 90 minutes. However, if a signal is observed upon revisiting the same position, it strongly implies a local type particle event. Otherwise, the particles should have moved elsewhere and could not be detected in the same location.

\begin{figure*}
\centering
\includegraphics[width=\textwidth]{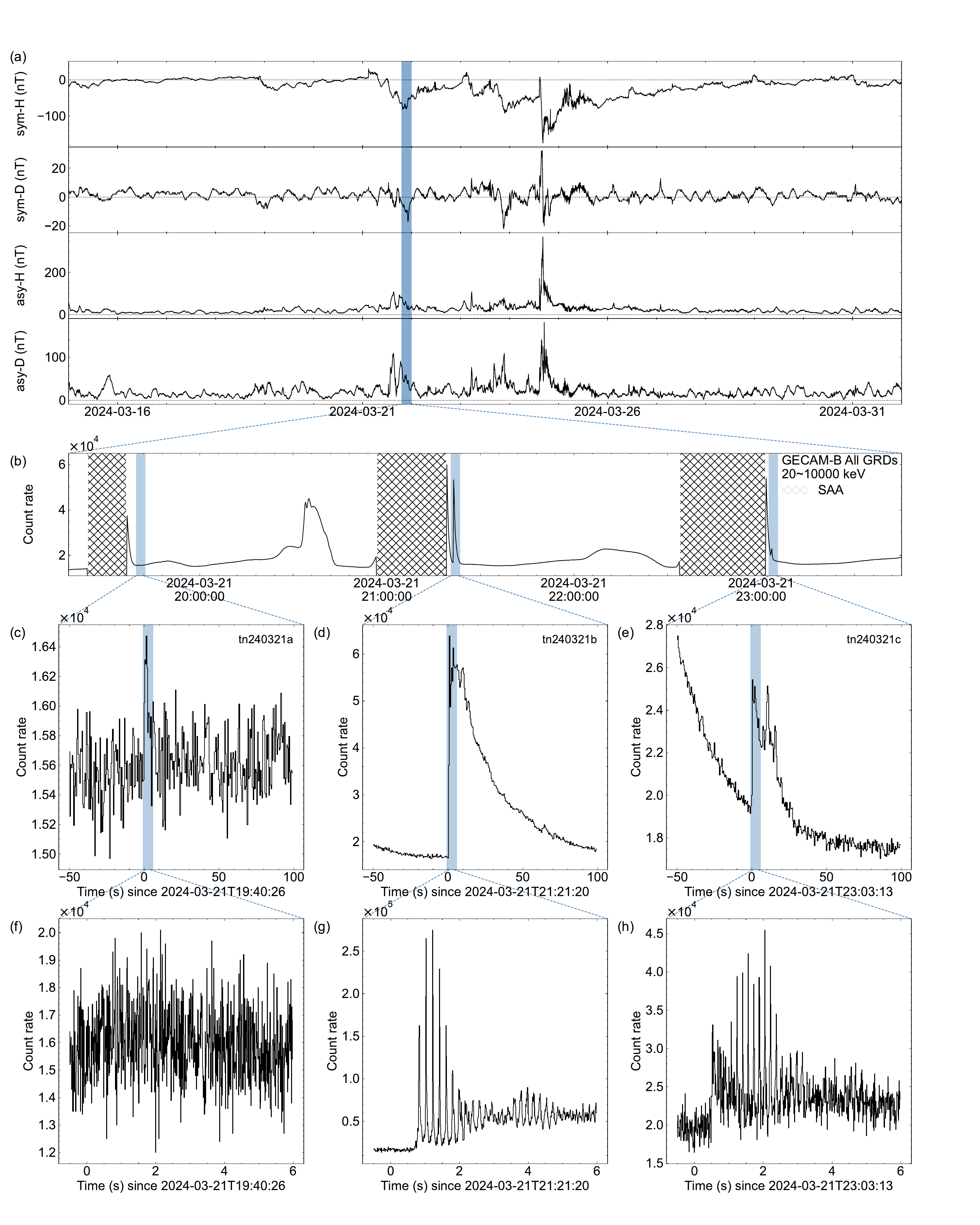}
\caption{\noindent\textbf{Overview of the count rate curve of OPP events in March 21, 2024.} 
a, the SYM/ASY index for Earth magnetic field, which is accessed from World Data Center for Geomagnetism, Kyoto. 
b, the count rate curve of GECAM-B of continuous 3 orbits with time resolution of 10\,s, the hatched time intervals corresponds to SAA, where GECAM-B is not operating. 
c, the count rate curve of tn240321a, tn240321b and tn240321c with time resolution of 500\,ms. 
d, the count rate curve of the peak time interval with time resolution of 10\,ms. For tn240321b and tn240321c, significant oscillation signals can be observed.}
\label{fig:240321_lc}
\end{figure*}

\begin{figure*}
\centering
\includegraphics[width=0.8\textwidth]{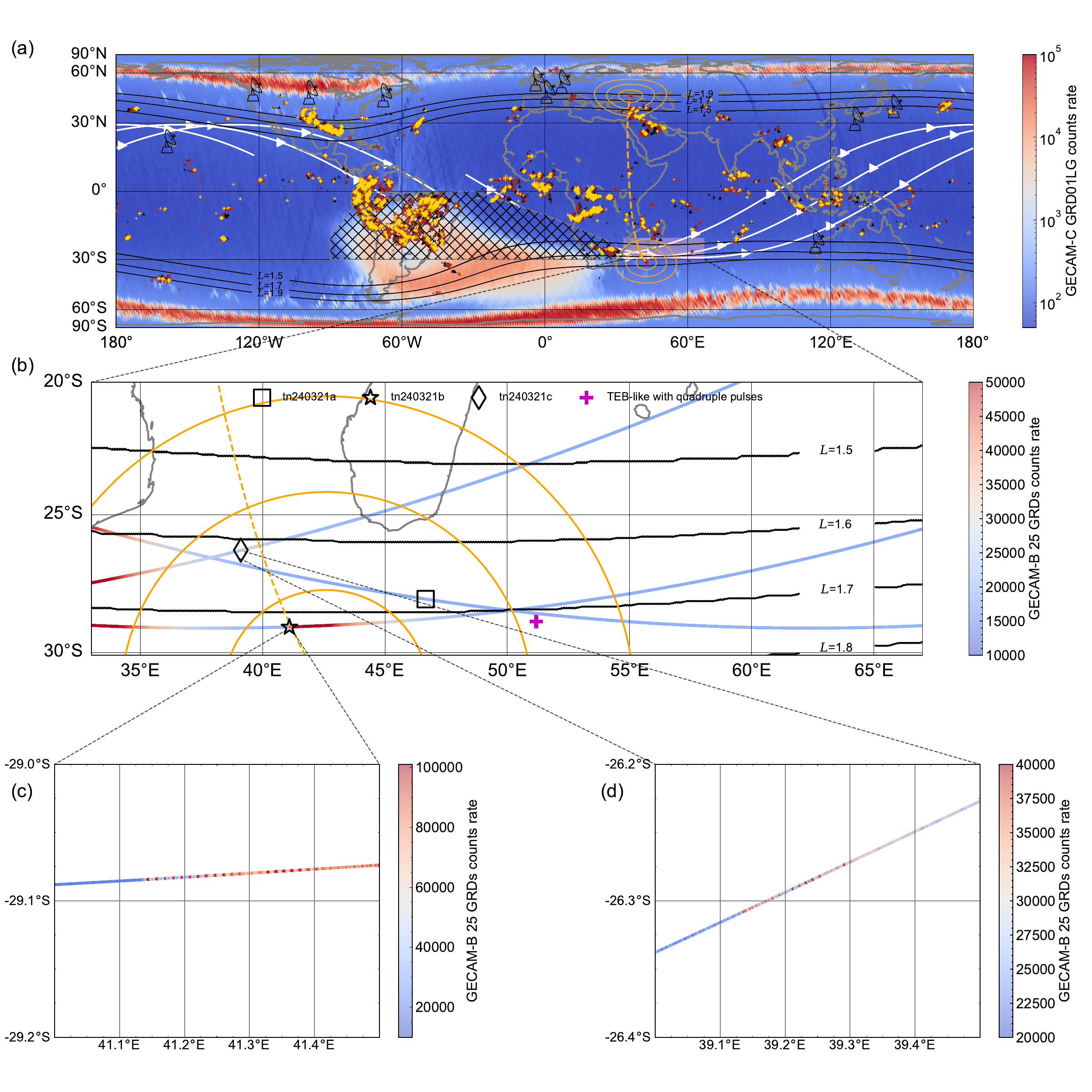}
\caption{\noindent\textbf{Overview of the map of OPP events in March 21, 2024.} 
a, the orbit of GECAM-B from 2024-03-21T19:00:00 to 2024-03-22T00:00:00 are depicted as white lines, and the white right triangle indicate the direction of movement of GECAM-B, with the time interval of 600\,s. The black triangle markers represent lightning from 2024-03-21T19:40:26 (trigger time of tn240321a) to 2024-03-21T21:21:20 (trigger time of tn240321b), while the brown cross markers are lightning from 2024-03-21T21:21:20 to 2024-03-21T23:03:13 (trigger time of tn240321c) and the gold plus markers shows the lightning after 2024-03-21T23:03:13 to the end of that day. The background color is obtained from the data of GECAM-C, representing the space environment at different locations. The black hashed part is defined as SAA for GECAM-B, and GECAM-B will not conduct observations in this area. The black lines represent different L-shell for GECAM-B orbit. The geomegnetic line tracing result of tn240321a are plotted as orange dashed line, while the magnetic footpoint are shown as red point. The three orange circles represent distances of 400, 800 and 1200km from the magnetic foot point, respectively. The locations of all VLF stations are also marked on the map.
b, the location of tn240321a and the twin OPP events, tn240321b and tn240321c are depicted as black square, pentagon and diamond. The magnta cross represent the location of the TEB-like events with quadruple pulses detected by GECAM-B at 2021-09-11T18:34:40. All these events occur near the south of Madagascar Island. It is interesting that the twin OPP events, tn240321b and tn240321c pass through the same magnetic field lines.
c and d, count rate of GECAM-B at each position. }
\label{fig:240321_loc}
\end{figure*}

Thanks to all the advantages of GECAM-B mentioned above, a series of interesting events are recorded on March 21, 2024, which was not a peaceful day for the magnetosphere of Earth. 
Accompanied by a series of Solar Flares (SFs) and Coronal Mass Ejections (CMEs) during the 25th solar activity cycle, sustained geomagnetic storms have occurred from March 21, which can be seen in the geomagnetic activity index shown in Fig.\ref{fig:240321_lc}\,a. 
An overview of the count rate curve of GECAM data from 2024-03-21T19:00:00 (UTC) to 2024-03-22T00:00:00 (UTC) is depicted in Fig.\ref{fig:240321_lc}\,b. 
GECAM-B was triggered \cite{Zhao2024trigger} by three interesting events at about 2024-03-21T19:40:26 (UTC), 2024-03-21T21:21:20 (UTC) and 2024-03-21T23:03:13 (UTC) when the satellite passed the south of Madagascar, which can be seen in Fig.\ref{fig:240321_loc}, and these three events are denoted as tn240321a, tn240321b and tn240321c respectively with the basic information listed in Table.\ref{tab:event_list}. 

As shown in Fig.\ref{fig:240321_lc}\,cde, the count rate curves of these events exhibit a high similarity to the typical shape of Gamma-ray bursts(GRBs) \cite{2018GRBbook}, one of the most violent explosive events in the universe, at low time resolution (0.5\,s), which usually has a fast rise and exponential decay (FRED) profile \cite{Norris_2005}. While for most of the charged particle events observed by GECAM and other wide-FOV gamma-ray monitors, their count rate curves show a slow rising behavior \cite{cai2025search}, and only very few events, such as TEBs, exhibit a rapid rising behavior, with a total duration of about tens of milliseconds. Therefore, a sudden increase with a timescale shorter than 1 second in charged particle events that last for a relatively long duration is unprecedented. 
However, due to the fact that CPDs also detected significantly strong signals at the same time, and no high energy transient signals were detected from the sky (including Solar) during tn240321b and tn240321c by analyzing the data of other X-ray and $\gamma$-ray monitors, these signals are undoubtedly dominated by charged particles rather than $\gamma$-ray produced by GRBs or other astrophysical events. 

Among these three events, the flux of tn240321b and tn240321c is much higher than that of tn240321a. Most importantly, as depicted in Fig.\ref{fig:240321_lc}\,g and h, high temporal resolution data reveal that there are very significant oscillation features around the peak stage. The rapid rising phase of tn240321b actually consists of intense pulses modulated periodically at more than 5 cycles per second. For each pulse, the rise time scale is on the order of milliseconds, which is significantly different from the slow evolution typically observed in ordinary particle precipitation events. 

To check the reliability of the signal and exclude instrument effects, we compared the engineering data and scientific data of tn240321b and tn240321c, both showing good consistency. And the peak count rate of GECAM-B for each detector is less than 20\,kcps, which is far below the saturation condition. Hence, instrument effects were ruled out as the origin of this oscillation signal, confirming it as OPP events with a physical origin. The continuity of time and consistency of location suggests their association, making them twin OPP events.

\section{Data Analysis and Discussion}\label{Sec_Dis}

The most charming feature of these three events is the strong quasi-periodic oscillation of tn240321b and tn240321c. 
The analysis includes aspects of temporal and spectral (taking the hardness ratio as an indicator). 
In this section, we will first discuss the detailed analysis results of tn240321a, tn240321b and tn240321c. 
Then we make efforts to identify the origin of this event.

\subsection{Temporal and Spectral Analysis}

\begin{figure*}
\centering
\begin{tabular}{cc}
\begin{overpic}[width=0.45\textwidth]{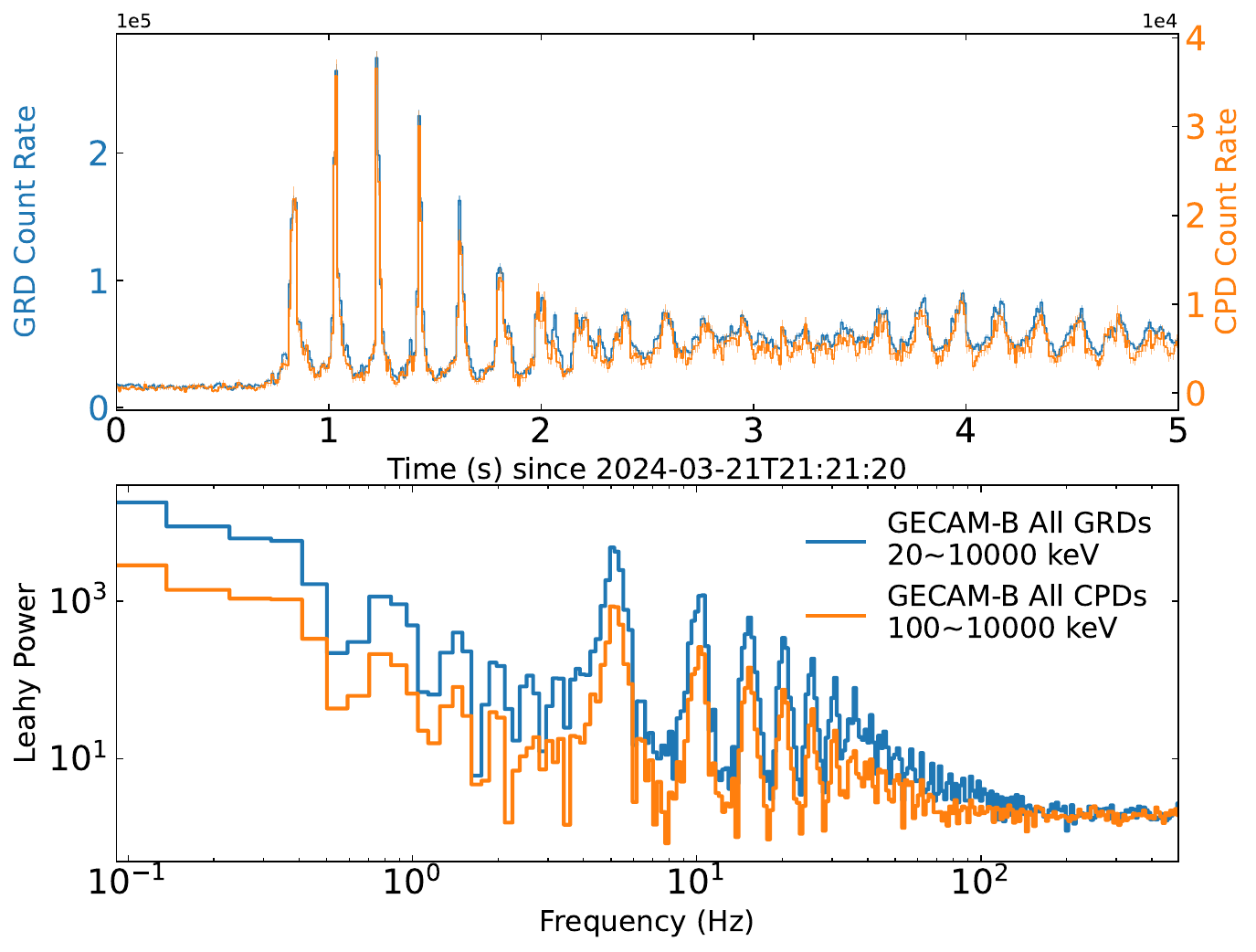}\put(-3, 70){\bf a}\end{overpic} &
        \begin{overpic}[width=0.45\textwidth]{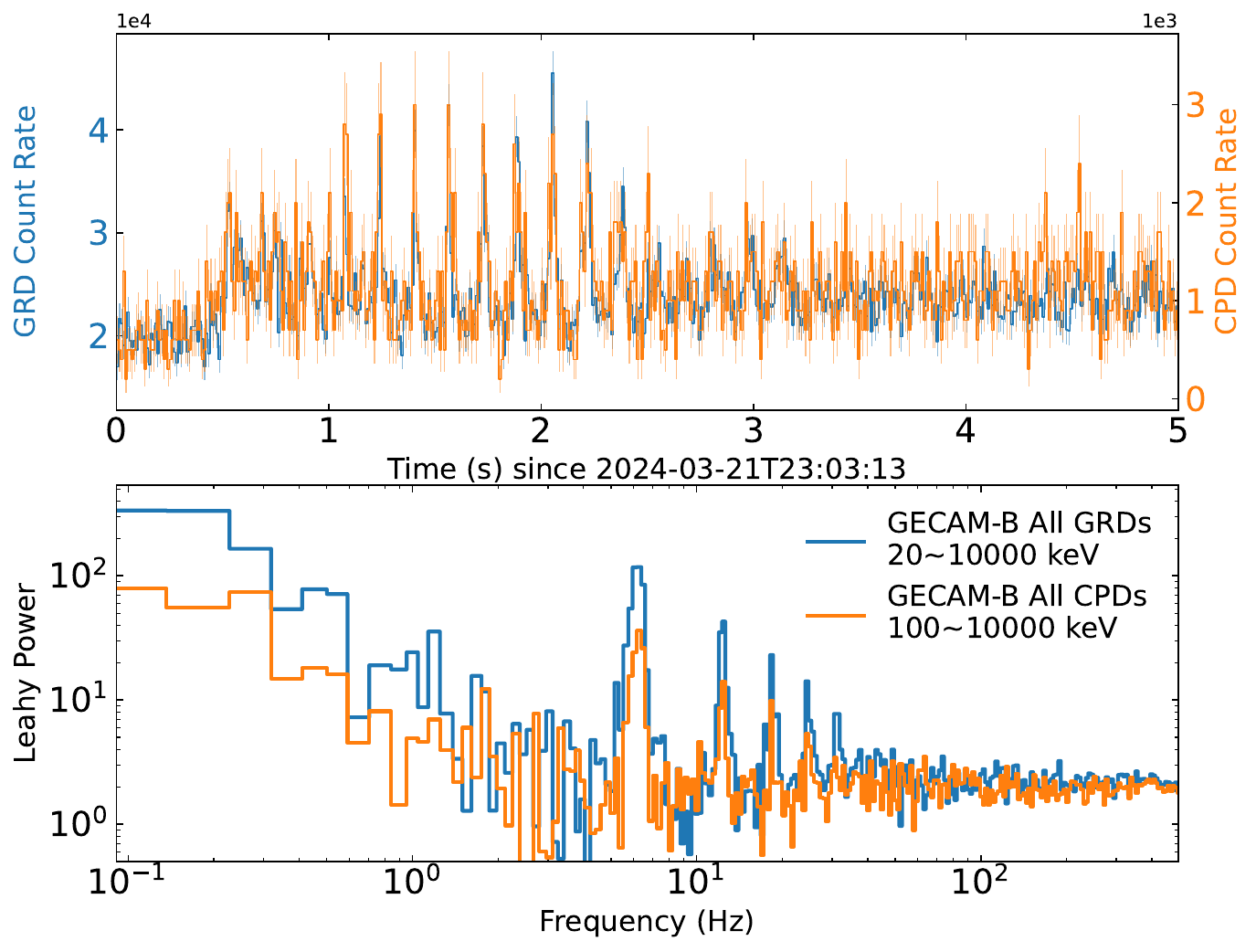}\put(-3, 70){\bf b}\end{overpic} \\
\begin{overpic}[width=0.45\textwidth]{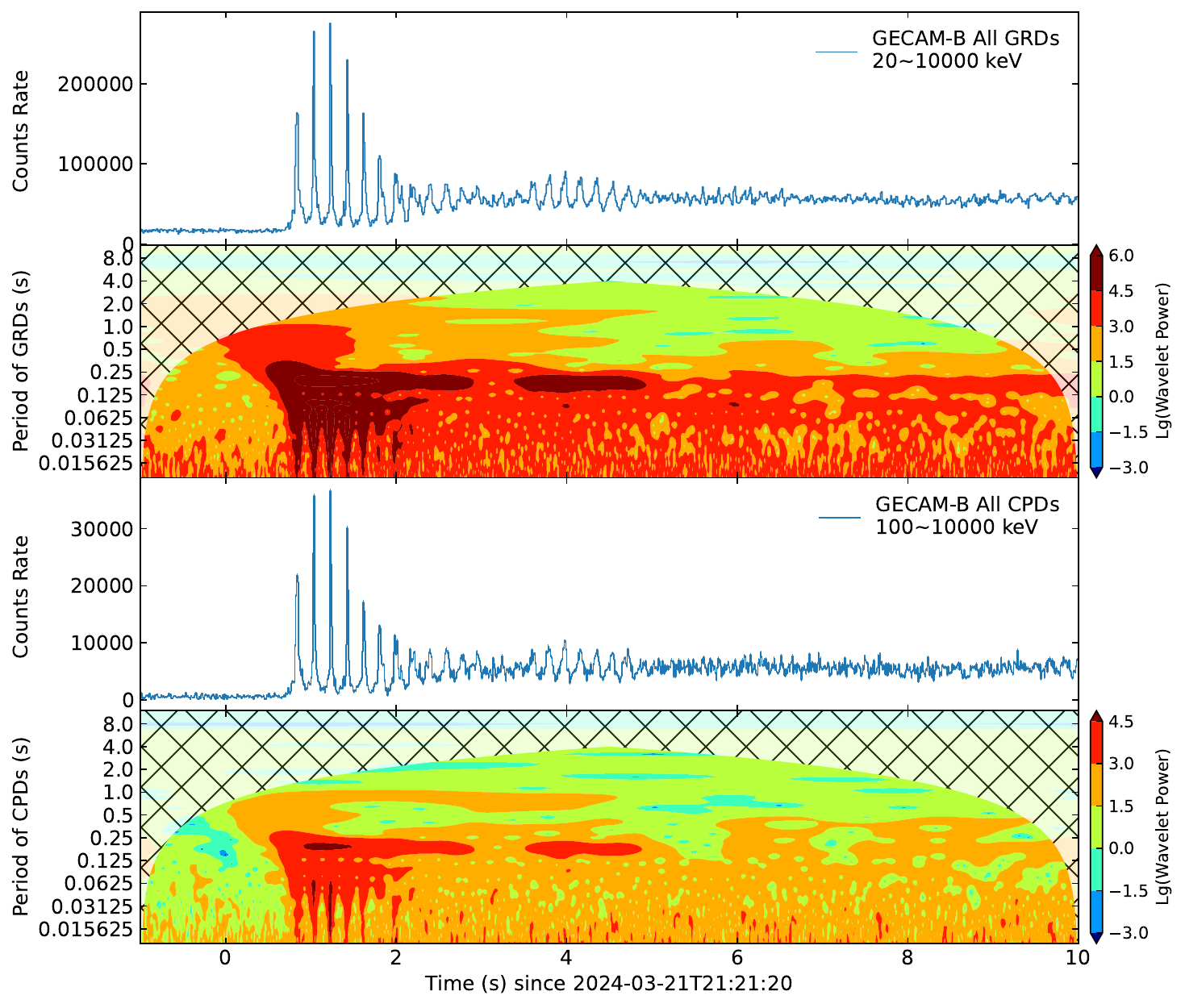}\put(-3, 80){\bf c}\end{overpic} &
        \begin{overpic}[width=0.45\textwidth]{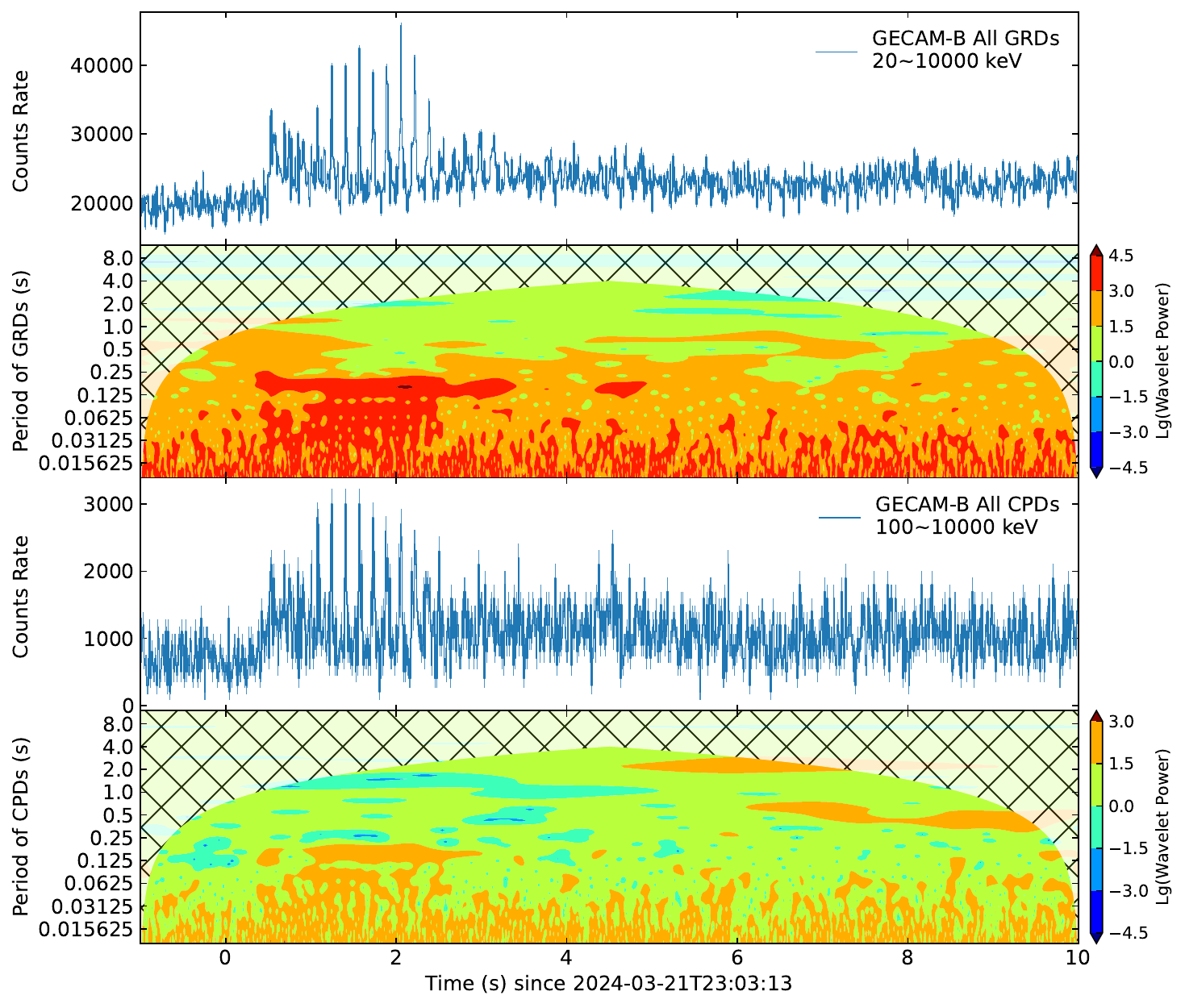}\put(-3, 80){\bf d}\end{overpic} \\
\begin{overpic}[width=0.45\textwidth]{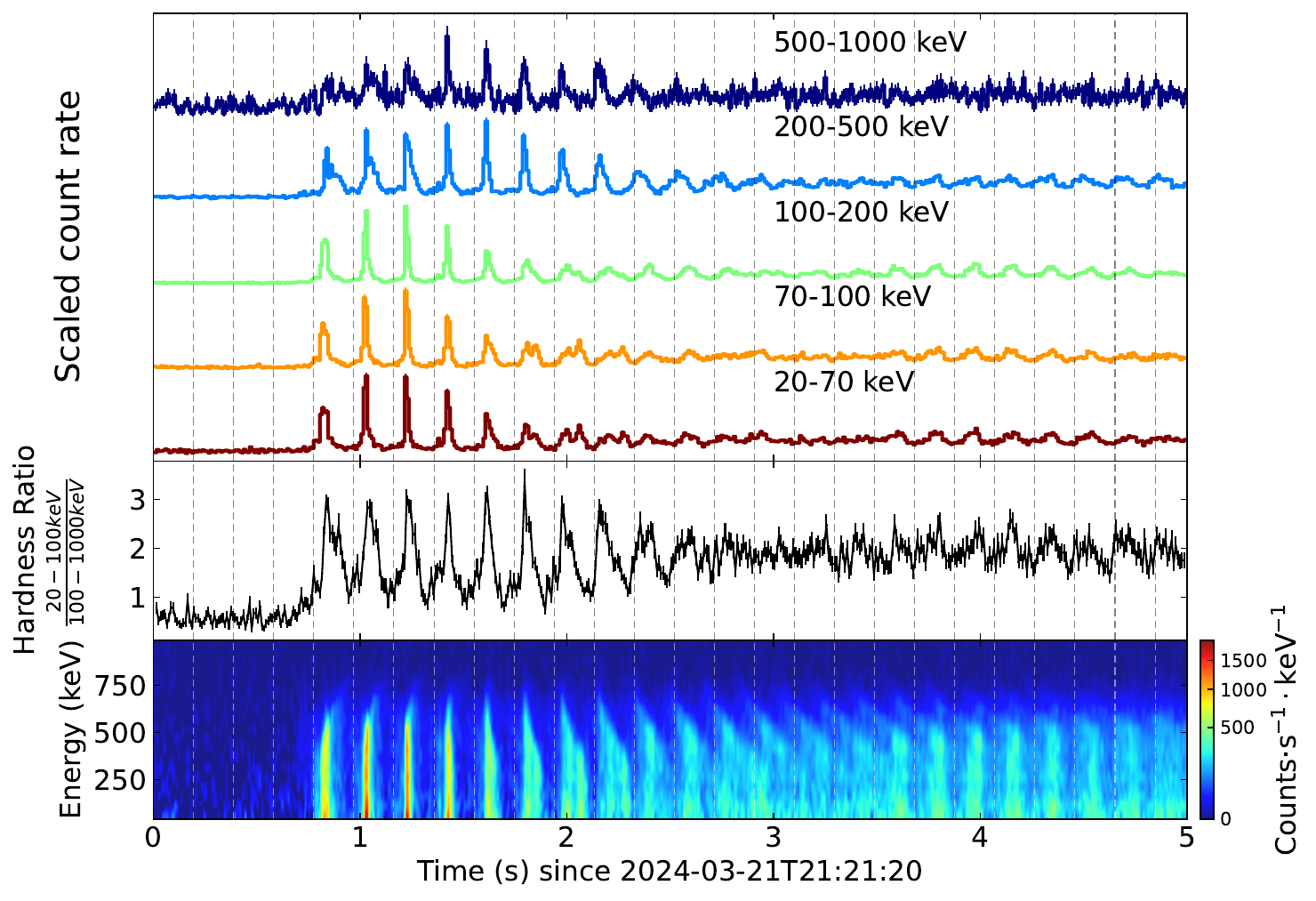}\put(-3, 70){\bf e}\end{overpic} &
        \begin{overpic}[width=0.45\textwidth]{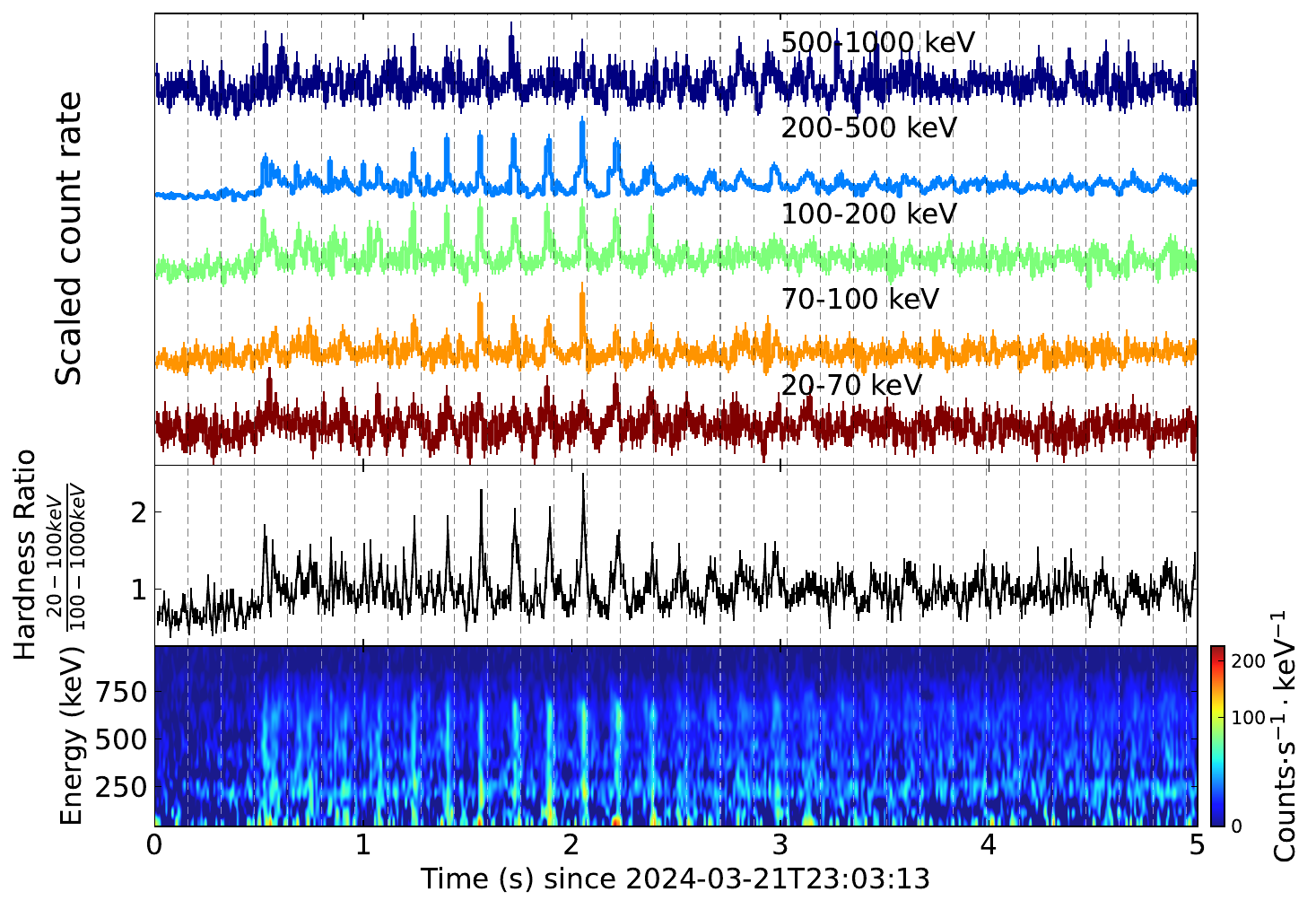}\put(-3, 70){\bf f}\end{overpic} \\
\end{tabular}
\caption{\noindent\textbf{The temporal analysis result from GRDs and CPDs of tn240321c and tn240321d.}
a and b, the top panels are count rate curve with time resolution of 10\,ms. The blue lines represent the data from GRDs while the data of orange lines are from CPDs. The bottom panel are PDS which is normalized by Leahy normalization. A series of harmonic and subharmonic can be clearly observed in the PSD of both GRDs and CPDs. The fundamental frequency of tn240321a is $\sim$5\,Hz while the fundamental frequency of tn240321b is $\sim$6\,Hz. 
c and d, the wavelet results of tn240321b indicate the presence of a frequency of approximately 5\,Hz for a long duration although the strength of the signal evolves over time. There is a brief period around 2024-03-21T21:21:23, when the oscillatory behavior almost disappears, but then becomes significant again. While for the tn240321c, the lifetime of the oscillatory behavior is much shorter. 
Both the frequency of tn240321b and tn240321c did not show any obvious evolution although the frequency are slightly different. 
e and f, the top panels are the scaled counts rate curve of multi-energy range from GRDs. For most events, oscillations are mainly observed below 500 keV, but there are also some events that exhibit significant oscillations in higher energy ranges. 
The middle panels are the curve of hardness ratio, which is calculated by the ratio of counts between 20 to 100 keV and 100 to 1000 keV. For time interval of high flux, the oscillations can also be clearly observed in hardness ratio. 
The bottom panel is the distribution of particles in energy-time in OPP events. 
We note that some pulses show the behavior of spectral lag and for some pulses exhibit ``softer-earlier" while some pulses exhibit ``softer-later". 
}
\label{fig:temporal}
\end{figure*}

The Leahy-normalized PSD of tn240321b and tn240321c, which are shown in Fig.\ref{fig:temporal}\,a and b respectively, is firstly calculated by \textit{Stingray} \cite{stingray_1,stingray_2} from the count rate curves of GRDs and CPDs with a time resolution of 1 ms. A series of harmonic and subharmonic frequencies can be clearly observed in the PSD. And we noticed that the frequencies of tn240321b and tn240321c are slightly different, with tn240321b having a fundamental frequency of approximately 5\,Hz, while tn240321c has a higher fundamental frequency of around 6\,Hz. 

Furthermore, we investigate the evolution of oscillation frequency over time by utilizing wavelet, as depicted in Fig.\ref{fig:temporal} c and d. The wavelet analysis of tn240321b confirms the presence of an oscillation frequency of approximately 5 Hz that persists for a long duration. Despite the oscillation intensity varying over time, we can observe that within a brief time interval around 2024-03-21T21:21:23, the oscillatory behavior nearly disappears, but subsequently becomes significant again. While for tn240321c, the duration of the oscillatory behavior is much shorter. Wavelet also indicates that tn240321c exhibits oscillation from the very beginning of the event, although these features cannot be directly discerned from the count rate curve (Fig.\ref{fig:temporal}\,d). Both tn240321b and tn240321c do not show any significant time evolution of frequency. 

Additionally, we analyze the oscillation feature of different energy ranges and noticed that they are not entirely the same, which can be seen in Fig.\ref{fig:temporal}\,e and f. The top panels of Fig.\ref{fig:temporal}\,e and d indicate that the oscillation of tn240321b has a wider energy distribution while much narrower in tn240321c. The oscillation is still clearly visible in the count rate curves above 500\,keV as well as in the signals below 100\,keV for tn240321b, while it is only visible in the signals in 100 to 500\,keV. The differences between the behaviors of different energy ranges result in significant oscillations in the hardness ratio as well, as shown in the middle panels in Fig.\ref{fig:temporal}\,e and f, which suggest that the spectrum of the peak part of mini-pulses is harder and the dip part is softer. 

When focusing on the energy-time distribution map of the charged particles in each mini-pulse (the bottom panels of Fig.\ref{fig:temporal}\,e and f), it is interesting to find that different pulses have different spectral lag properties with an overall evolution, particularly for tn240321b. From the perspective of temporal features, the initial few pulses exhibit a ``softer-earlier" behavior (i.e. the low energy pulse arrives earlier than the high energy pulse). However, the low energy and high energy charged particles gradually turn to reach the peak at the same time,  and eventually transition to a ``softer-later" behavior in the later mini-pulses. 

Given the high speed of the satellite, these temporal oscillations likely indicate a periodic location distribution of particles, with spatial intervals of $\sim$1409 km (tn240321b) and $\sim$1134 km (tn240321c). 
% Considering that the satellite is moving fast, these temporal oscillations may actually suggest the periodic location distribution of particles, with tn240321b having a spatial interval of $\sim$1409 km and tn240321c a spatial interval of $\sim$1134 km. 
And from the perspective of spatial evolution, the detected spectral lag evolving over time in mini-pulses implies that the location distribution of high energy particles is narrower and more ``compact" than low energy charged particles. 

In addition to the spectral lag of each mini-pulse, the envelope profile of the overall tn240321b also exhibits a ``softer-earlier" spectral lag feature. We can see this clearly from the relative amplitude of the mini-pulses, as the second mini-pulse is the highest pulse among the pulses in the energy range of 20-70\,keV, while the initial few pulses have very low relative amplitude in the light curve at 500-1000\,keV.

Although tn240321a itself does not have many interesting features compared to tn240321b and tn240321c, it provides a wealth of reference information to help explore the origin of these events. The positions of these three events overall exhibit a trend of ``west to east". And if these three events are considered to be physically associated, then we can obtain a constraint on the duration of the particle precipitation, which lasted for at least 3.5 hours.

We also note that tn240321b and tn240321c occurred in the same region as the first multi-peak TEB-like event reported by \cite{YIZ_TGF_GEC_2023}, with a similar oscillation frequency. This implies that there may be some connection between OPP events and multi-peak TEB-like events, two phenomena with durations differing by two orders of magnitude. They may both be related to the complex geomagnetic field near Madagascar or some transient structures in the magnetosphere, such as magnetic holes \cite{NC_Li2020,APJL_Huang2017} although no magnetic holes have been detected in this region, which is still an open question.

\subsection{Possible Origin}
Overall, tn240321a, tn240321b, and tn240321c exhibit many properties that show similarities with various known phenomena, yet differ in certain aspects. Together with potentially related activities, including geomagnetic disturbances and lightning, we discuss the possible physical origins one by one.

\subsubsection{Artificial VLF Wave}
Due to the very significant and stable periodicity of these two OPP events, it is natural to try to associate them with human activities.
We explored the possibility that these events have an artificial origin. As is widely known, human activities, such as VLF, could potentially generate anomalous stripes in space. One of the most well-known artificial electron belts is the North-West Cape (NWC) electron belt, and it is already known that the NWC electron belt can be frequently observed by GECAM-B. 

Some major VLF transmitters are plotted in Fig.\ref{fig:240321_loc}(a) while no VLF transmitter and their magnetic mirror point is near the location of OPP events tn240321b and tn240321c. 
Considering the drift motion of electrons in the artificial electron belt, the position on the GECAM-B orbit with the same magnetic shell parameter \textit{L} as the NWC transmitter (\textit{L}$\sim$1.6) is also marked on the map, and this line does indeed pass through the region of tn240321b and tn240321c. However, the drift motion of these electrons is typically eastward, and the region of tn240321b and tn240321c is located to the west of the NWC transmitter. This means that if the local charged particles of tn240321a and tn240321b are caused by the electron drift of the NWC electron belt, the drifting electrons would need to travel almost a full orbit around the Earth, making this a low possibility. 
Additionally, the energy distribution of the OPP events is also different from the previous detection of the NWC belt by other instruments such as China Seismo-Electromagnetic Satellite (CSES) \cite{NWC_Li_2019} and DEMETER \cite{NWC_Zhang_2016}, in that the electron energy of the electrons in the NWC electron belt is typically lower than 300\,keV, much lower than the energy of charged particles observed in tn240321b and tn240321c, if the charged particles are electrons. 
Similar conclusions can be drawn for the transmitters in Europe. Additionally, because the magnetic mirror points of these transmitters stations are located within the SAA, charged particles tend to precipitate directly in the SAA, making it difficult for them to drift eastward.

\subsubsection{Geomagnetic Storm and Substorms}
A more likely origin of OPP is particle precipitation caused by geomagnetic activity. Since these three events occurred near the peak time of a geomagnetic storm, and location is very close to the SAA as shown in the Fig.\ref{fig:240321_loc}\,a. 
It is possible that some charged particles in SAA escaped due to the weakening of the magnetic field during the geomagnetic activity and then were captured by a temporary precipitation belt. 
It needs to be noted that the local solar time (LST) for these three events is all at midnight, which is the period when geomagnetic substorms frequently occur. However, due to the limited number of samples of OPP events, we cannot conclude whether OPP is actually associated with geomagnetic storms or substorms. 

\subsubsection{Microburst}
Another possible related phenomenon is microburst, which is first observed by \cite{GRL_Anderson1964} as the short X-ray pulses in the aurora zone and originates from the bremsstrahlung radiation of scattered precipitation electrons by chorus wave. Although the precipitation electron of microbursts shows the similar pulse width and energy distribution as OPP events \cite{GRL_Barcus1966,GRL_Oliven1968}, the location of the OPP events does not support this being a microburst, as the statistical result on the occurrence locations of microbursts is highly consistent with the aurora zone and nearly no detection in the low latitude region \cite{GRL_Douma2017}. Indeed, some microbursts observed near the equatorial plane were reported \cite{GRL_Shumko2018}, but this event occurred in Earth's outer radiation belt with a large magnetic shell parameter \textit{L} which is significantly distinguished from OPPs. Additionally, microbursts do not show strong periodic signals and no overall profile evolution, though they are composed of repetitive narrow pulses \cite{GRL_Kandar2023,GRL_Wetzel2024}, which is also different from OPPs. Hence, OPP should be a phenomenon different from microbursts.

\subsubsection{Lightning-induced Events}
There are two major types of charged particle events that can be induced by lightning: one is TEBs , and the other is LEPs. 
For the former, charged particles are directly produced by the lightning process, and the high-speed electron beam is generated through the avalanche multiplication \cite{Carlson_TEB_2009,Dwyer_TEB_2008,GBM_TEB}. While for the latter, charged particles trapped in the radiation belt are scattered into the loss cone via cyclotron resonance with the whistler-mode wave generated by lightning, leading to precipitations \cite{LEP_SPA_Voss1984}. Theoretically, both TEBs and LEPs can exhibit multi-peaks signals through multiple bounces of electrons when certain conditions are met. It has also been observed that lightning during geomagnetic storms may generate OPP-like events \cite{feinland2024lightning} with L-shell ($\sim2$) generally larger than tn240321b and tn240321c. 

However, TEB samples are very rare, and their durations are generally very short (on the order of tens of milliseconds), typically allowing for the detection of only a bounced pulse. Only one case of a suspected multiperiodic TEB event has been reported until now, but its origin is still under debate. As for tn240321b and tn240321c, although their period is similar to that of TEB-like with quadruple pulses on 2021-09-11T18:34:40, which can be seen in Figure\,4c of Zhao et al. \cite{YIZ_TGF_GEC_2023}, their duration is significantly longer than that of TEB or even TEB-like. 
Moreover, the WWLLN data indicates that there are no signs of lightning activity within 1200 km from the GECAM-B nadir and southern magnetic footpoint of tn40321a and tn40321b. Continuous lightning activity has been observed around the northern magnetic foot point, but it is all located more than 400 km away and is neither frequent nor concentrated (if lightning occurs densely within a small area, it may mistakenly result in multiple independent TEBs being interpreted as a multiphase charged particle event). Therefore, it is doubtful that tn240321b and tn240321c are caused by the lightnings directly in the form of TEB. The lack of strong association with lightning presents some challenges to the LEP scenario as well. 
% But we also note that the lightnings cluster around northern magnetic foot point is located at the east of the Mediterranean, where

Moreover, if the multi-pulses structure of tn240321a is attributed to multiple bounces of electrons, it is difficult to understand why mini-pulses exhibit a transition from ``softer-earlier" to ``softer-later". The energy evolution of the envelope profile of the count rate curve is also in conflict with the electrons bouncing scenario, as the first few mini-pulses exhibit a gradually increasing trend. 
Additionally, since the travel times of charged particles from satellites to the geomagnetic north and south foot points are different, it is doubtful that multiple bounces could generate pulse sequence with equal time intervals.

A possible alternative explanation is that charged particles of different energies are bound at different locations, and the satellites passing through these dense and sparse regions may lead to such oscillation signals in flux temporally, and the envelope curves of the count rates at different energies can have varying shapes and allow for the rise of mini-pulses. 
Therefore, whether tn240321a, tn240321b, and tn240321c represent a special case of LEP still remains to be analyzed in detail through modeling (in preparation).

\section{Summary and Conclusion}\label{Sec_Con}
Thanks to the delicate and innovative design of detectors and electronics, GECAM, originally designed for time-domain astronomy, has detected many space charged particle events with fast variability. 
Notably, on March 21, 2024, GECAM-B fortunately detected three interesting particle precipitation events (i.e. tn240321a, tn240321b and tn240321c), two of which (tn240321b and tn240321c) exhibited significant periodicity. 
Through rigorous analysis, we have excluded instrumental effect origins for the periodic signals and conclusively identified these phenomena as local space charged particle events (i.e. OPP events) with intrinsic oscillatory characteristics. 

The emphasis of this paper is on the new features revealed by the high resolution measurement from GECAM. Utilizing GECAM-B GRDs and CPDs data, several interesting temporal and spectral features of these particle precipitation events were observed for the first time:
\begin{enumerate}[(1)]
\item The oscillation, as depicted by base frequency of two OPP events are different (i.e. $\sim$5\,Hz for tn240321b and $\sim$6\,Hz for tn240321c), and the frequency does not show obvious evolution over time for each event, although the possibility that tn240321b and tn240321c together follow a slow evolution of frequency with $\dot{f}\sim0.2\,\rm mHz/s$ cannot be ruled out. 
\item The amplitude of the oscillations varies among different energy ranges, resulting in the same modulation in the hardness ratio. 
\item The oscillations in different energy ranges also exhibit significant spectral lag, including spectral lag of narrow pulses and the profile of the overall envelope. 
\end{enumerate}

The spectral features and location provided evidence that suggested that OPPs may be related to magnetic field activity but do not support tn240321a, tn240321b and tn240321c as an artificiality from VLF stations. 
The comparison between the twin OPP events on March 21, 2024, and the multi-peak TEB-like event on September 9, 2021, indicates that the origins of their periodicity may be similar. 
But the association with lightning is not favored as there is no supporting evidence by analysis time coincidence and localization coincidence between OPP events and lightning activity with WWLLN data. 
And these findings challenge the conventional opinion that attributes the oscillations to electron bounce motion. 

The high time resolution observations and report of OPPs here by GECAM demonstrate the huge potential of GECAM in space physics research, especially in events with extremely high flux or rapidly changing temporal structures. 
Moreover, these phenomena also suggest that particle precipitation events exhibit more mysterious and complex behaviors, providing profound insights into understanding the space environment.

Given their unprecedented features, the origin of these three OPP events still remains an enigma. They could be either a new type of particle precipitation event or a kind of peculiar LEP. 
Fortunately, in the past few years of operation, GECAM has observed a series of OPP events as well as multi-peak TEB-like events. The systematically compiled catalog (in preparation) obtained through comprehensive searches will enable statistical classification and investigation of these events. 
Additionally, some other time-domain astronomy telescopes such as \textit{Insight}-HXMT \cite{HXM_INS_Zhang2020} and SVOM/GRM \cite{SVM_GRM_Dong2010} are expected to be able to detect OPP events as well. 
Joint observation with space detectors and ground instruments will help reveal the origin of this peculiar oscillation phenomenon in feature as well.

\Acknowledgements{This work is supported by 
the National Natural Science Foundation of China 
(Grant No. 12273042, % xiong shaolin 
12303045), % caice 
the National Key R\&D Program of China (Grant No. 2021YFA0718500), % xiong shaolin for HXMT & GECAM
the Natural Science Foundation of Hebei Province (No. A2023205020) % cai ce
and the Strategic Priority Research Program of the Chinese Academy of Sciences (Grant No. 
XDB0550300, % xiong shaolin, Tao lian for xiandaoB
XDA30050000). % xiong shaolin for GTM
The GECAM (Huairou-1) mission is supported by the Strategic Priority Research Program on Space Science of the Chinese Academy of Sciences (XDA15360000). 
We thank the World Wide Lightning Location Network (http://wwlln.net) as a collaboration of more than 50 universities. 
Chenwei Wang appreciate Wen Cheng, Yixin Sun, Lu Wang for helpful discussions.
}

\InterestConflict{The authors declare that they have no conflict of interest.}

\begin{table*}[htbp]
\scriptsize
\caption{\centering{Basic information of three events in March 21, 2024}}
\begin{tabular*}{\textwidth}{@{}ccccccccccc@{}}
% \toprule
\hline
 & & & & \multicolumn{3}{c}{Nadir of GECAM-B} & \multicolumn{2}{c}{Southern magnetic footpoint$^1$} & \multicolumn{2}{c}{Northern magnetic footpoint$^1$} \\
% \hline
\cmidrule(r){1-4} \cmidrule(r){5-7} \cmidrule(r){8-9} \cmidrule(r){10-11}
ID & Trigger UTC Time & LST & SYM-H & Longitude & Latitude & Altitude & Longitude & Latitude & Longitude & Latitude \\
  &   & (hh:mm:ss) & (nT) & (Deg) & (Deg) & (km) & (Deg) & (Deg) & (Deg) & (Deg)\\
\hline
tn240321a & 2024-03-21T19:40:26 & 22:40:26 & -78 & 46.66 & -28.09 & 597 & 48.52 & -30.49 & 42.05 & 43.26\\
tn240321b & 2024-03-21T21:21:20 & 00:21:20 & -74 & 41.09 & -29.09 & 594 & 42.59 & -31.37 & 35.99 & 44.17\\
tn240321c & 2024-03-21T23:03:13 & 02:03:13 & -53 & 39.10 & -26.32 & 597 & 40.42 & -28.74 & 35.34 & 42.06\\
\hline
\end{tabular*}
\label{tab:event_list}
\begin{tablenotes}
\item[1] $^1$ The longitude and latitude of the southern and northern magnetic footpoint are set as altitude of 40 km.
\end{tablenotes}
\end{table*}

\newpage

\bibliographystyle{unsrt}
\bibliography{reference}

\end{multicols}
\end{document}